\documentclass{IET-Conf-Paper}

\usepackage{lscape}
\usepackage{comment}

\usepackage{amsmath,amsbsy}
\usepackage{mathdots}
\usepackage{yhmath}
\usepackage{siunitx}
\usepackage{array}
\usepackage{gensymb}
\usepackage{amssymb}
\usepackage{mathtools}     

\usepackage{amsmath,esint}

\usepackage{bm}
\usepackage{latexsym}
\usepackage{amssymb}
\usepackage{amsmath}
\usepackage{amsfonts}
\usepackage{siunitx}
\usepackage{lscape}
\usepackage{float}

\usepackage[hyphens]{url}
\usepackage{xcolor,colortbl}

\definecolor{Gray1}{gray}{0.85}
\definecolor{Gray2}{gray}{0.93}
\definecolor{LightCyan}{rgb}{0.88,1,1}
\newcolumntype{a}{>{\columncolor{Gray}}c}
\RequirePackage{booktabs}

\begin{document}

\title{POWER-SWING BLOCKING SCHEME BASED ON CONTINUOUS AND HIGHER-ORDER INCREMENTAL QUANTITIES FOR SYSTEMS WITH INVERTER-BASED RESOURCES \vspace{0.5 cm}}

\author{Henrik Johansson\textsuperscript{\textsc{\textit{1}}}*, Qianli Xing\textsuperscript{\textsc{\textit{1}}}, Nathaniel Taylor\textsuperscript{\textsc{\textit{1}}}, Xiongfei Wang\textsuperscript{\textsc{\textit{2}}} }

\address{\textsuperscript{\textsc{\textit{1}}}{Department of Electric Power and Energy Systems, KTH Royal Institute of Technology, Stockholm, Sweden}\\
\textsuperscript{\textsc{\textit{2}}}{Department of Electrical Engineering, Tsinghua University, Beijing, China}
\\ *E-mail: hjo5@kth.se}



\keywords{POWER-SWING BLOCKING, INVERTER-BASED RESOURCES, INCREMENTAL QUANTITIES, OUT-OF-STEP TRIPPING}

\begin{abstract}
Conventional power-swing blocking (PSB) schemes for distance protection face new and emerging challenges as the penetration levels of inverter-based resources (IBRs) increase. Several papers highlight interoperability issues concerning faster power swings and unconventional impedance trajectories. Revised relay settings are proposed in attempts to deal with the problems. However, the blinder-based PSB methodology suffers from being dependent on the speed of the apparent impedance trajectory over the tiny region between the blinders, making it difficult to distinguish between fast swings and faults. Moreover, blinders have the inherent limitation that they cannot operate if the measured impedance is inside the blinders when the fault is cleared and immediately enters the distance zones during the first swing cycle---an issue often overlooked in the current literature. This paper proposes a novel PSB scheme based on higher-order incremental quantities (IQs) for difficult swing scenarios with IBRs and synchronous generators (SGs). PSCAD/EMTDC simulations of an SG and an implemented grid-forming (GFM) inverter validate its superior performance compared to conventional dual-blinder PSB schemes. The impacts of GFM inverters on the impedance measured by distance relays during the fault-to-swing transition and first power swing cycle are also discussed.

\end{abstract}

\maketitle

\section{Introduction} 
Transmission line faults, loss of substantial generation or load, line switching and other large power system disturbances cause the system to embark on a power swing where the angles of the sources oscillate. During power swings, the measured loop impedances of distance relays may enter their trip zones, causing unwanted tripping and risks cascading into a blackout. To prevent this from happening, a power-swing blocking (PSB) scheme accompanies distance relays \cite{PSB-Tutorial-Normann}. Its role is to detect power swing conditions and block the distance relay from tripping in such cases, without impacting the distance relay dependability during faults. In case loss of synchronism occurs, the out-of-step tripping (OST) function should subsequently detect the unstable power swing condition in order to execute a controlled separation of the system at predetermined generation-load balance locations to avoid equipment damage from overvoltages and overcurrents, and to prevent a widespread power outage through cascaded contingency tripping \cite{PSB-OOS-fundamentals}. Today, dual-blinder impedance-based schemes are commonly used for PSB and OST in conventional distance relays \cite{PSB-Bogdan-Zcont, REL670-application-manual, D60-instruction-manual-GE, SEL-421-4, CIGRE-Protection-Challenges-IBRs}. These schemes assume that the measured apparent positive-sequence impedance $\bar{Z}_1$ quickly crosses the blinders during faults but significantly slower during power swings. 

Due to the continuous increase of inverter-based resources (IBRs) in modern power systems, such as wind and solar power, several recent papers have studied the impact of both grid-following (GFL) and grid-forming (GFM) inverters on the conventional dual-blinder PSB/OST methodology \cite{PSB-impact-of-wind, PSB-IBRs-PartII, PSB-GFM-VI, PSB-Yongxin-GFL, PSB-Yongxin-GFM, PSB-GFM-Protection-and-Dynamics}. These papers conclude that the PSB and OST settings should be revised, mainly concerning faster power swings and $\bar{Z}_1$ trajectories differing from those of synchronous generators (SGs). This includes adjusting the PSB pickup time, OST logic and blinder zones. However, faster PSB assertion can cause unwanted blocking of distance relays during actual faults and compromise the system transient stability. Therefore, distance relay manufacturers tend to advise against using a pickup threshold lower than 1.5 fundamental periods \cite{PSB-IBRs-PartII}, at least for the first PSB activation. For example, in \cite{REL670-application-manual}, the PSB pickup times for the initial and subsequent swing cycles in a $50$~Hz power system are calculated to $30$ and $10$~ms, respectively. 

From the current state-of-the-art literature, it is seen that the traditional dual-blinder PSB methodology has limitations considering IBRs. Moreover, blinders have the inherent limitation that they cannot operate if $\bar{Z}_1$ is inside the blinders when the fault is cleared and immediately enters the distance trip zones---an issue often overlooked in the current literature with IBRs \cite{PSB-impact-of-wind, PSB-IBRs-PartII, PSB-GFM-VI, PSB-Yongxin-GFL, PSB-Yongxin-GFM, PSB-GFM-Protection-and-Dynamics}. Hence, it is important to investigate alternative ways to realize the PSB function in distance relays for inverter-dominated grids. In this paper, a novel PSB scheme based on continuous and higher-order incremental quantities (IQs) is proposed. It reliably blocks the distance relay during fast $\bar{Z}_1$ power swing trajectories produced from IBRs. The higher-order IQs are used to achieve quick PSB activation during the fault-to-swing transition of the power system, while still allowing the backup distance zones to operate in case of long-lasting faults that also continuously produce IQs. Therefore, the proposed PSB scheme successfully prevents unwanted tripping of the distance relay even if $\bar{Z}_1$ immediately enters distance zone 1 when the fault is cleared. In conventional SG-based power systems, situations like these may occur due to unstable or marginally stable power swings caused by long-lasting faults. However, this paper shows that such scenarios are more prone to occur with GFM inverters, owing to their high-magnitude and dynamic source impedances during power swings. The proposed IQ-based PSB scheme and a traditional dual-blinder PSB scheme are implemented in MATLAB (an OST scheme also accompanies them). They are evaluated and compared on extensive PSCAD/EMTDC simulation data with an SG and GFM inverter implementation. Python is used to automate the PSCAD simulations. 

Following this introduction, Section.~\ref{sec:System-description} describes the power system model, the distance relay with a blinder-based PSB/OST scheme, and the GFM inverter model implemented in PSCAD. Section.~\ref{sec:initial-power-oscillation} explains characteristics of the initial power oscillation with SGs and GFM inverters, as well as limitations of blinders for PSB. Section~\ref{sec:higher-order-IQs} discusses higher orders of IQs and proposes an IQ-based PSB scheme. Section~\ref{sec:simulation-results} compares the performance of the proposed IQ-based PSB scheme with that of a conventional blinder-based scheme on PSCAD simulations. Section~\ref{sec:conclusions} concludes the findings of the paper.

\section{System Description}
\label{sec:System-description}
\begin{figure}[h]
    \centering
    \includegraphics[width=0.9\linewidth]{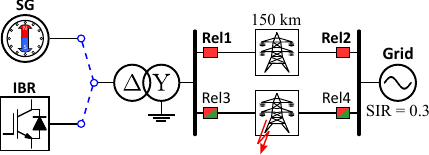}
    \caption{Benchmark two-source parallel line system to generate power swings.}
    \label{fig:two-source-parallel-ohl}
\end{figure}

The system shown in Fig.~\ref{fig:two-source-parallel-ohl} is implemented in PSCAD to generate power swings. The sending-end source (left) is either an SG ($H=3$~s) or an IBR, and the remote-end source (right) is the utility grid with a source-to-line impedance ratio (SIR) of either $0.3$, $0.6$ or $0.9$ for each line. The analyzed quadrilateral distance relays are ``Rel1'' and ``Rel2'', which have a reach of $80\%$, $135\%$ and $200\%$ of the total line length for their zone 1, 2 and 3, respectively. Zone 1, 2 and 3 trip within 5, 450 and 1 000 ms, respectively. The resistive reach extensions have been tuned to accurately detect faults when an SG is connected at the sending end (see \cite{HJ-GFM-distance} for more details). To cause power swings, balanced faults are applied along the lower line and cleared after a certain time. Both lines are modelled using the frequency-dependent (phase) model available in PSCAD. The system voltage, frequency and nominal power are 220~kV, 50~Hz and 300~MVA, respectively. The line is 150~km and the simulated balanced fault-to-swing cases with either an SG or GFM inverter in Fig.~\ref{fig:two-source-parallel-ohl} are:
\begin{itemize}
    \item Distance from the relay $m_\mathrm{f}\in \{10, 30, 75, 120, 140 \}$~km 
    \item Fault resistance $R_\mathrm{f} \in \{0, 1, 5, 10, 15\}$ $\Omega$
    \item Fault clearing time \newline $t_\mathrm{c} \in \{40, 60, 100, 150, 200, 300, 400, 500, 10000\}$~ms 
\end{itemize}

\subsection{Implemented Distance Relay and Blinders}
\begin{figure}[h]
    \centering
    \includegraphics[width=0.65\linewidth]{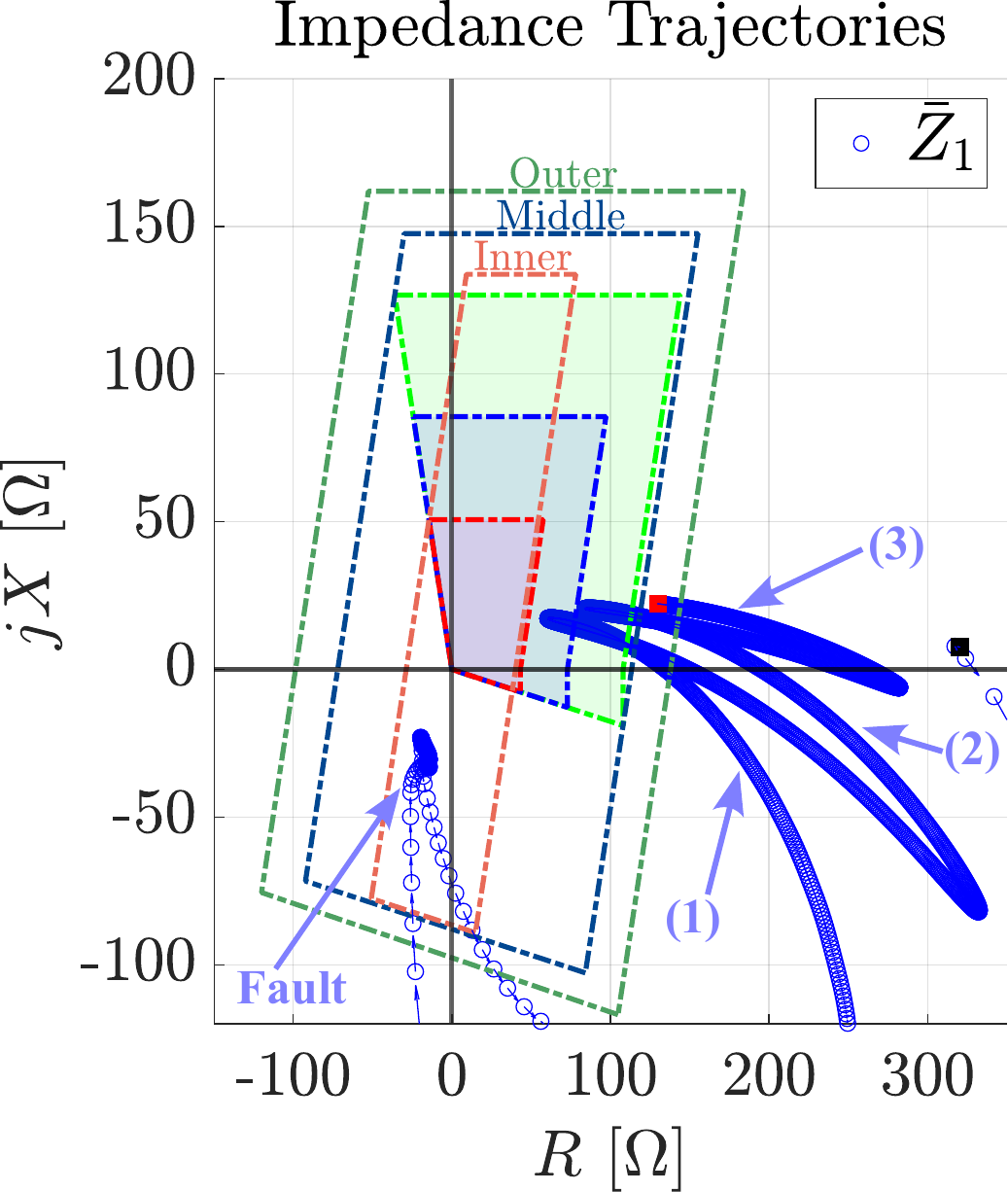}
    \caption{Distance relay model and dual-blinder PSB/OST zones.}
    \label{fig:Blinder-example}
\end{figure}

The quadrilateral distance relay and the concentric blinder zones used for the blinder-based PSB/OST scheme are shown in Fig.~\ref{fig:Blinder-example}. A stable SG-based power swing observed in Rel1 after a close-in fault on the parallel line in Fig.~\ref{fig:two-source-parallel-ohl} is illustrated, where the time between two samples of $\bar{Z}_1$ is $1$~ms, the black mark is the initial load point, the red mark is the final sample considered in the plot, and the numbers indicate the chronological swing cycles. If $\bar{Z}_1$ takes longer than $\Delta T_\mathrm{PSB,1}=30$~ms to cross between the outer and middle blinders, a PSB signal is issued, and for subsequent swing cycles, the PSB pickup time is $\Delta T_\mathrm{PSB,2}=10$~ms. Otherwise, the distance zones are allowed to operate as usual. The right outer blinder has been designed to not infringe on the load region of Rel1 after the disconnection of one line. The inner blinder is used to detect loss of synchronism; if $\bar{Z}_1$ enters the inner blinder two times while the PSB is active, an OST signal is issued. The reason why the implemented OST requires $\bar{Z}_1$ to enter the innermost blinder two times is because droop-based GFM inverters can resynchronize despite completing a full angular rotation if a stable active power operating point exists, thanks to their first-order synchronization dynamics, which should not be mistaken for an unstable swing. Section~\ref{sec:simulation-results} will show such a stable power swing. Through extensive transient stability analysis when an SG is the sending-end source, the right inner blinder is tuned so that the most severe stable swings (marginally stable) observed in Rel1 do not cross it. The reverse reach has been set according to proposed blinder design guidelines with GFM inverters \cite{PSB-Yongxin-GFM}.

\subsection{Implemented GFM Inverter Control Strategy}

\begin{figure}[h]
    \centering
    \includegraphics[width=1\linewidth]{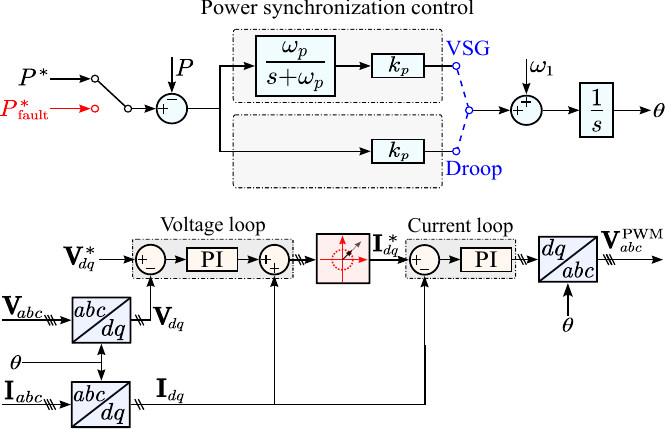}
    \caption{Control diagram of the implemented GFM inverter.}
    \label{fig:GFM1-vsg-droop}
\end{figure}

The GFM inverter control diagram is shown in Fig.~\ref{fig:GFM1-vsg-droop}. It employs either virtual synchronous generator (VSG) or droop-based power synchronization control (PSC), and uses a $1.2$~pu magnitude saturation block for current limitation without any angle adjustment. This current limiter provides accurate and fast current limitation for GFM inverters during power swings; however, it emulates an adaptive virtual resistance, jeopardizing the otherwise homogeneous system impedance angles in a predominantly inductive power system \cite{GFM-limiter-sat-resistive}. When the point of common coupling (PCC) voltage collapses, for example during faults or power swings, the active power reference $P^*$ is reduced. More control details can be found in \cite{HJ-GFM-distance}.

\section{Initial Power Swing Oscillation}
\label{sec:initial-power-oscillation}
Previous works have found that IBRs tend to cause faster power swings and unconventional trajectories of $\bar{Z}_1$ compared to SGs \cite{PSB-impact-of-wind, PSB-IBRs-PartII, PSB-GFM-VI, PSB-Yongxin-GFL, PSB-Yongxin-GFM, PSB-GFM-Protection-and-Dynamics}. However, the impacts of IBRs on $\bar{Z}_1$ during the fault-to-swing transition period and initial power swing oscillation cycle are often neglected. In this section, the initial power oscillation will be explained with SGs and GFM inverters to highlight limitations of blinder-based PSB schemes.

\subsection{Limitations of Blinders in SG-dominated Systems}
\label{sec:SG-swings-explained}
\begin{figure}[h]
    \centering
    \includegraphics[width=0.75\linewidth]{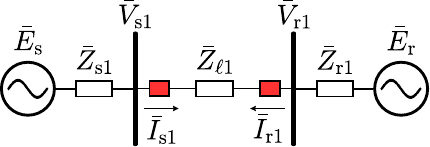}
    \caption{Two-source equivalent system.}
    \label{fig:two-source-swing}
\end{figure}

When the faulted line is cleared in Fig.~\ref{fig:two-source-parallel-ohl}, the two-source equivalent system in Fig.~\ref{fig:two-source-swing} is obtained. The sending- and remote-end voltage sources have the internal voltages $\bar{E}_\mathrm{s} = E_\mathrm{s}\angle\delta_\mathrm{s}$ and $\bar{E}_\mathrm{r} = E_\mathrm{r}\angle\delta_\mathrm{r}$, respectively. The sending-end relay (subscript ``s'') computes the apparent impedance 
\begin{equation}
\begin{aligned}
    \bar{Z}_1 &= \frac{\bar{V}_\mathrm{s1}}{\bar{I}_\mathrm{s1}} =\frac{\bar{E}_\mathrm{s}-\bar{I}_\mathrm{s1}\bar{Z}_\mathrm{s1}}{\bar{I}_\mathrm{s1}} = \frac{\bar{E}_\mathrm{s}}{\bar{E}_\mathrm{s}-\bar{E}_\mathrm{r}}\bar{Z}_\mathrm{sr\ell1}-\bar{Z}_\mathrm{s1},
\end{aligned} 
\label{Z1-apparent-general}
\end{equation}
where $\bar{Z}_\mathrm{sr\ell1} = \bar{Z}_\mathrm{s1}+\bar{Z}_\mathrm{r1}+\bar{Z}_\mathrm{\ell1}$. A time-varying angle difference $\delta(t) = \delta_\mathrm{s}(t)-\delta_\mathrm{r}(t)$ between the two equivalent sources yields a moving apparent impedance $\bar{Z}_1(t)$. Its trajectory may be over simple or complicated paths, as well as having a near constant slip angular velocity $d\delta/dt=\omega_\mathrm{slip} = 2\pi f_\mathrm{slip}$ or any arbitrary function of time $\delta_\mathrm{s}'(t)-\delta_\mathrm{r}'(t)$.

When the fault is cleared in Fig.~\ref{fig:two-source-parallel-ohl}, the system in Fig.~\ref{fig:two-source-swing} has an initial value of $\delta$. This value, together with the system impedances $\bar{Z}_\mathrm{s1}$, $\bar{Z}_\mathrm{r1}$ and $\bar{Z}_\mathrm{\ell1}$, dictate the starting value of $\bar{Z}_1$ during the initial power oscillation. Due to the sliding-window discrete Fourier transform (DFT), it takes about one fundamental period for $\bar{Z}_1$ to transition from its fault value to its initial power swing value, and the path it takes is the fault-to-swing transition trajectory. 

In Fig.~\ref{fig:Blinder-example}, the $\bar{Z}_1$ fault-to-swing transition trajectory goes outside the blinders, allowing the blinder-based PSB scheme to pick up on the initial power oscillation. However, for other values of $\delta$ and/or system impedances, $\bar{Z}_1$ may not leave the blinders during the fault-to-swing transition period and immediately enter the distance zones. Blinders cannot operate for such power swings, risking unwanted tripping of the remaining healthy line, particularly by distance zones 1 and 2. Moreover, if the fault on the lower line in Fig.~\ref{fig:two-source-parallel-ohl} is located close to the utility grid, Rel1 sees a zone 2 fault, which may lead to $\bar{Z}_1$ quickly entering distance zone 1 during the fault-to-swing transition period. These problematic scenarios can occur under both stable and unstable power swings, depending on the angular acceleration $\delta''(t)$ and its velocity $\delta'(t)$ reached when the fault is cleared at the initial angle $\delta$. With SGs, long-lasting faults tend to be more problematic in this regard since $\delta(t)$ has more time to advance from the load region. Hence, marginally stable or unstable power swings are more prone to suffer from these initial power oscillations. 

For the case when an SG is connected to the utility grid in Fig.~\ref{fig:two-source-parallel-ohl}, an electrical angle of $\delta=180^\circ$ implies that the machine slips a magnetic pole, typically making the two portions of the power system in Fig.~\ref{fig:two-source-swing} enter an out-of-step (OOS) condition and lose synchronism. When $\delta=180^\circ$, $\bar{Z}_1$ reaches the so-called electrical center (EC). Its location in the impedance plane depends on the system impedances $\bar{Z}_\mathrm{s1}$, $\bar{Z}_\mathrm{\ell1}$ and $\bar{Z}_\mathrm{r1}$, as well as the voltage magnitudes $E_\mathrm{s}$ and $E_\mathrm{r}$. If $E_\mathrm{s}=E_\mathrm{r}$, it holds that \cite{PSB-impact-of-wind}
\begin{equation}
    \bar{Z}_1 = \frac{\bar{Z}_\mathrm{sr\ell1}}{2} \Bigg( 1-j\cot\bigg(\frac{\delta}{2} \bigg) \Bigg)-\bar{Z}_\mathrm{s1}
    \label{Z1-delta},
\end{equation}
where the trajectory of $\bar{Z}_1$ follows along a straight line that intersects $\bar{Z}_\mathrm{sr\ell1}$ orthogonally at its middle point. When $\bar{Z}_\mathrm{s1}=\bar{Z}_\mathrm{r1}$, the EC is located at the middle of the transmission line, i.e. at $0.5\bar{Z}_\mathrm{\ell 1}$. If $E_\mathrm{s}\neq E_\mathrm{r}$, the EC shifts toward the lower voltage source and $\bar{Z}_1$ travels along a circle \cite{PSB-Tutorial-Normann}.

\subsection{Impact of GFM Inverters on Power Swings and the Initial Power Oscillation}
\label{sec:GFM-swings-explained}
\begin{figure}[h]
    \centering
    \includegraphics[width=1\linewidth]{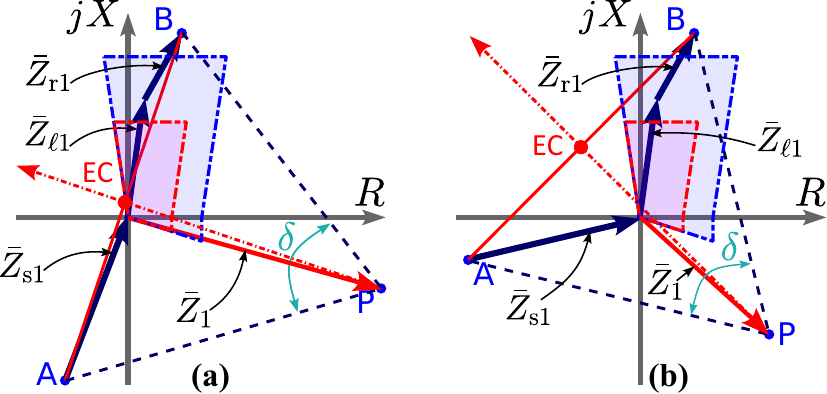}
    \caption{EC location and $\bar{Z}_1$ trajectory outlook from (\ref{Z1-delta}) for two momentary values of $\bar{Z}_\mathrm{s1}$: \textbf{(a)} inductive $\bar{Z}_\mathrm{s1}$; \textbf{(b)} resistive $\bar{Z}_\mathrm{s1}$.}
    \label{fig:Z-traj-EC}
\end{figure}

The source impedance $\bar{Z}_\mathrm{s1}$ of GFM inverters increases during faults to limit the injected current. During power swings, the current limiter is engaged for certain intervals of $\delta$ to limit the current by virtue of a dynamical source impedance. Both its magnitude and angle may change during power swings, depending on the current limiter of the GFM inverter. A varying $\bar{Z}_\mathrm{s1}$ causes the EC location to vary as function of $\delta$, impacting the trajectory of $\bar{Z}_1$ and its value when $\delta=180^\circ$. For example, the EC is pushed downwards if $|\bar{Z}_\mathrm{s1}|\gg |\bar{Z}_\mathrm{r1}+\bar{Z}_\mathrm{\ell1}|$ and $\angle\bar{Z}_\mathrm{s1}\approx \angle\bar{Z}_\mathrm{r1} \approx \angle\bar{Z}_\mathrm{\ell1}$, as would be the case with a GFM inverter employing a predominantly inductive virtual impedance for current limitation. Fig.~\ref{fig:Z-traj-EC}~(a) shows this situation for a certain value of $\delta$. As $\delta$ increases towards $180^\circ$, $|\bar{Z}_\mathrm{s1}|$ needs to increase in order to limit the injected current to tolerable values, which eventually moves the EC below the distance zones and cause $\bar{Z}_1$ to travel below the distance zones. If the GFM inverter instead emulates a predominantly resistive virtual impedance in $\bar{Z}_\mathrm{s1}$ for current limitation, as the GFM inverter in Fig.~\ref{fig:GFM1-vsg-droop}, $\bar{Z}_1$ may travel below and then behind the distance zones because the EC is pushed sideways into the second quadrant as $\delta$ increases towards $180^\circ$. Fig.~\ref{fig:Z-traj-EC}~(b) illustrates this situation. These $\bar{Z}_1$ trajectories have been shown with GFM inverters \cite{PSB-GFM-VI, PSB-Yongxin-GFM}.

Because GFM inverters push the EC location below and behind the distance zones, there is a high risk that $\bar{Z}_1$ quickly enters distance zone 1 during the fault-to-swing transition trajectory after zone 2 faults are cleared on the parallel line in Fig.~\ref{fig:two-source-parallel-ohl}. This can happen over a wide range of clearing times because $|\bar{Z}_\mathrm{s1}|$ needs to be significantly increased to limit the injected current over a wide range of $\delta$. In Fig.~\ref{fig:Z-traj-EC}~(a), $\bar{Z}_1$ would transition from its zone 2 fault value to point ``P'' when the fault is cleared at the illustrated $\delta$. The produced $\bar{Z}_1$ fault-to-swing transition trajectory may cross distance zone 1, and the risk increases as the clearing angle $\delta$ gets closer to $180^\circ$. The likelihood that the fault-to-swing transition trajectory of $\bar{Z}_1$ crosses distance zone 1 in Fig.~\ref{fig:Z-traj-EC}~(b) after a zone 2 fault is even greater, because point ``P'' is almost right below the distance zones even for this relatively low clearing angle $\delta$. As the clearing angle increases up to $180^\circ$, point ``P'' moves left towards the third quadrant and then up into the second quadrant. For all these values of $\delta$, there is a high risk that $\bar{Z}_1$ quickly enters distance zone 1 and cause unwanted tripping during the fault-to-swing transition trajectory after zone 2 faults are cleared.

\section{Higher Orders of IQs and a Proposed IQ-Based PSB Scheme}
\label{sec:higher-order-IQs}
Since IBRs tend to cause faster power swings, unconventional $\bar{Z}_1$ trajectories compared to SGs and increase the likelihood of causing $\bar{Z}_1$ fault-to-swing transition trajectories that blinder-based PSB schemes cannot detect, it is important to investigate alternative PSB methods that are less impacted by these power swing characteristics. In this section, such a PSB scheme is proposed based on continuous and higher-order IQs. 

\subsection{Different Orders of IQs}
During power swings, voltages and currents vary over long time periods. The IQs represent these changes by being the difference between the present data samples and the samples stored in memory $p$ fundamental periods back in time:
\begin{equation}
    \begin{aligned}
        &\Delta v_{x}(t) = v_{x}(t) - v_{x}(t-pT) \\
        &\Delta i_{x}(t) = i_{x}(t) - i_{x}(t-pT),
    \end{aligned}
    \label{IQ-definition}
\end{equation}
where $x\in \{\mathrm{a, b, c}\}$ is any phase and $T$ is the fundamental time period (20~ms in $50$~Hz systems). These characteristics of IQs are different compared those during line faults, which cause the IQs to get a significant, sudden and short-lived increase in magnitude. While (\ref{IQ-definition}) is the typical definition of IQs, different orders of IQs may be considered. For example, the first three orders of incremental currents may be defined according to
\begin{equation}
    \begin{aligned}
        &\Delta i_{x}'(t) = i_{x}(t) - i_{x}(t-pT) \\
        & \Delta i_x''(t) = i_{x}(t) - 2i_{x}(t-pT) + i_{x}(t-2pT) \\
        &\Delta i_x'''(t) = i_x(t) - 3i_x(t-pT) + 3i_x(t-2pT) - i_x(t-3pT),
    \end{aligned}
    \label{IQ-current-orders}
\end{equation}
where the different orders of IQs resemble the numerical approximations of the first, second and third derivatives. 

The order of incremental phase currents yields different characteristics of the incremental positive-sequence current magnitude during power swings: $|\Delta \bar{I}_1'(t)|$, $|\Delta \bar{I}_1''(t)|$ and $|\Delta \bar{I}_1'''(t)|$. For example, a continuous linear increase in the amplitude of $i_x(t)$ will energize and increase $|\Delta \bar{I}_1'(t)|$, but $|\Delta \bar{I}_1''(t)|$ and $|\Delta \bar{I}_1'''(t)|$ remain close to zero. A continuous quadratic increase in the amplitude of $i_x(t)$ energizes both $|\Delta \bar{I}_1'(t)|$ and $|\Delta \bar{I}_1''(t)|$ in different amounts, but $|\Delta \bar{I}_1'''(t)|$ still remains close to zero. Hence, the IQ-order $n$ has a high rejection rate against lower order system dynamics. When transmission line faults occur, the order of IQs does not matter under the first $p$ post-fault cycles if a stable pre-fault load condition existed.

\subsection{Proposed IQ-Based PSB Scheme}

\begin{figure}[h]
    \centering
    \includegraphics[width=0.7\linewidth]{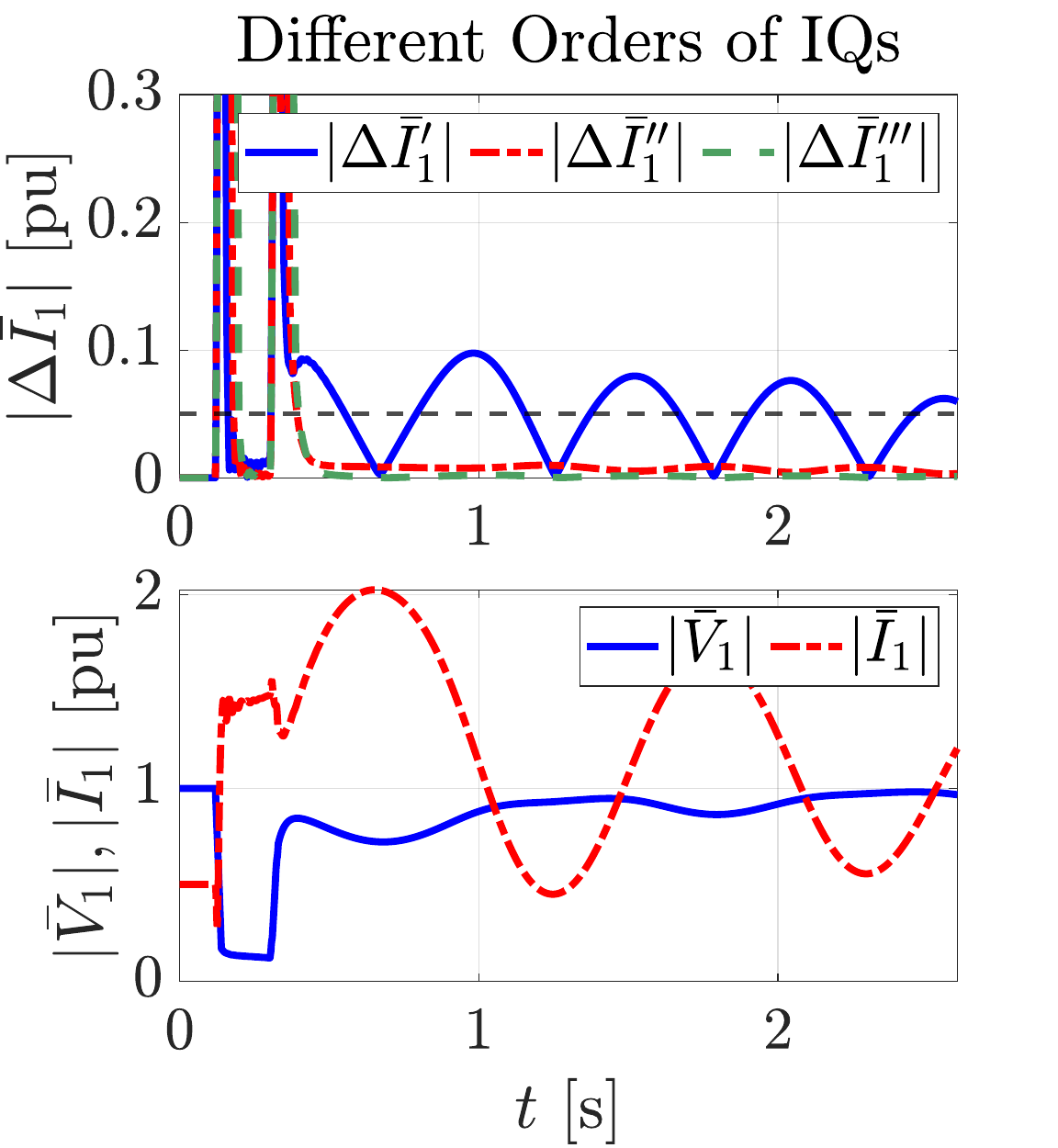}
    \caption{Different orders of IQs from a stable power swing.}
    \label{fig:PSB-IQ-example-SG}
\end{figure}

Continuous IQs for PSB is discussed in \cite{PSB-Tutorial-Normann}. A PSB signal is issued if the incremental current is greater than a threshold of $\Delta I_\mathrm{th} = 0.05$~pu for $\Delta T_\mathrm{PSB,IQ} = 60$~ms continuously (three cycles). The method uses the first-order incremental quantities and may operate on a loop basis or use the positive-sequence incremental current magnitude $|\Delta\bar{I}_1'|$, which averages out the balanced oscillations occurring in each phase from the sliding-window DFT during power swings. However, there are several issues with this simple implementation:
\begin{itemize}
    \item Unwanted PSB activation may happen during long-lasting faults that produce continuous $|\Delta\bar{I}_1'|$, blocking backup distance protection. This depends on $\Delta I_\mathrm{th}$, $\Delta T_\mathrm{PSB,IQ}$, the fault location, fault resistance and angular dynamics of the sources. A low threshold level $\Delta I_\mathrm{th} = 0.05$~pu and pickup time $\Delta T_\mathrm{PSB,IQ} = 60$~ms are unlikely to restrain for a wide range of possible faults. In general, as $\bar{Z}_1$ moves around in the impedance plane during faults, incremental currents are also produced and must be accounted for. 
    
    \item Increasing $\Delta I_\mathrm{th}$ for security purposes during faults may result in failure to activate when needed during power swings, particularly slow swing frequencies. 
    
    \item The three-cycles pickup time $\Delta T_\mathrm{PSB,IQ} = 60$~ms is too slow to prevent the healthy line in Fig.~\ref{fig:two-source-parallel-ohl} from tripping if $\bar{Z}_1$ enters distance zone 1 during the fault-to-swing transition trajectory. On the other hand, a faster pickup time is practically infeasible because, even with $p=1$, $|\Delta\bar{I}_1'|$ from (\ref{IQ-current-orders}) will typically remain above $\Delta I_\mathrm{th}$ for a minimum of 40~ms (two cycles) when a fault occurs, due to the delay coming with the sliding-window DFT. This time can also become significantly longer than 40~ms if $\bar{Z}_1$ is moving around in the impedance plane. 

    \item $|\Delta\bar{I}_1'|$ becomes close to zero when the rate of change of the positive-sequence voltage $\bar{V}_1$ and current $\bar{I}_1$ are almost zero. This is problematic because when $\bar{V}_1$ is at a local minimum and $\bar{I}_1$ at a local maximum, $\bar{Z}_1$ measured by the distance relay is low in magnitude and has the highest possibility of being inside the distance zones during the swing cycle. This is often the time when PSB is needed. The issue is illustrated for a stable power swing from an SG in Fig.~\ref{fig:PSB-IQ-example-SG}. The second- and third-order IQs are also included. The horizontal grey line is $\Delta I_\mathrm{th}=0.05$~pu. 
\end{itemize}

To overcome the above mentioned design issues, the proposed IQ-based PSB scheme also utilizes higher-order IQs. As can be seen in Fig.~\ref{fig:PSB-IQ-example-SG}, the higher-order IQs $|\Delta\bar{I}_1''|$ and $|\Delta\bar{I}_1'''|$ remain well below the $\Delta I_\mathrm{th} = 0.05$~pu threshold even during the power swing itself. Hence, compared to $|\Delta\bar{I}_1'|$, they are also considerably less likely to be above $\Delta I_\mathrm{th}$ longer than expected during faults if $\bar{Z}_1$ moves around in the impedance plane. If $p=1$, it is expected that $|\Delta\bar{I}_1''|$ and $|\Delta\bar{I}_1'''|$ remain above $\Delta I_\mathrm{th}$ for a minimum of 60~ms (three cycles) and 80~ms (four cycles) during faults, respectively, due to the delay caused by the sliding-window DFT. When the fault is cleared, both $|\Delta\bar{I}_1''|$ and $|\Delta\bar{I}_1'''|$ once again experience sharp and sudden increases in their values, as seen in Fig.~\ref{fig:PSB-IQ-example-SG}. This information can be used by distance relays to estimate when the fault is cleared and the system enters a power swing condition. In essence, a separate PSB activation function needs to first be triggered before the main PSB function based on continuous values of $|\Delta\bar{I}_1'|$ is enabled.

The proposed PSB activation function is as follows:
\begin{enumerate}
    \item When the fault first occurs, $|\Delta\bar{I}_1''|$ shoots above $\Delta I_\mathrm{th}$. This moment is stored as an estimate of the fault inception time.
    \item $|\Delta\bar{I}_1''|$ later falls below $\Delta I_\mathrm{th}$, after which a PSB activation signal is put on hold. 
    \item The PSB activation signal is triggered when $|\Delta\bar{I}_1''|>\Delta I_\mathrm{th}$ for the second time. 
    \item Once triggered, the PSB signal is maintained as long as $\bar{Z}_1$ is inside a supervisory zone covering the distance zones, here chosen as the middle concentric blinders in Fig.~\ref{fig:Blinder-example}. This ensures the PSB function blocks the distance relay immediately after detecting the fault clearance if $|\Delta\bar{I}_1'|$ is accumulating and has not yet continuously been above $\Delta I_\mathrm{th}$ longer than the pickup time $\Delta T_\mathrm{PSB,IQ}$. If instead $|\Delta\bar{I}_1'|$ has been above $\Delta I_\mathrm{th}$ longer than the pickup time before the fault is cleared, the PSB activation function prevents it from blocking the distance zones during the fault. 
\end{enumerate}
When computing the IQs, $p=1$ is used in (\ref{IQ-current-orders}). $|\Delta\bar{I}_1''|$ is used for the PSB activation function instead of $|\Delta\bar{I}_1'''|$ because $|\Delta\bar{I}_1''|$ settles below $\Delta I_\mathrm{th}$ faster after the fault occurs, allowing the PSB activation function to be ready for the fault clearance faster. However, there is a risk that the PSB activation function spuriously advances in its steps due to noise or short-lived oscillations on $|\Delta\bar{I}_1''|$ during its transient caused by the fault, particularly in step 2 above when the transient expires. To prevent this, a $2T=40$~ms time delay is implemented between counting each crossing of $|\Delta\bar{I}_1''|$ over $\Delta I_\mathrm{th}$, or rather when $|\Delta\bar{I}_1''|-\Delta I_\mathrm{th}$ changes sign. 

After the PSB activation function has been triggered, the main PSB function takes over, which activates if $|\Delta\bar{I}_1'|>\Delta I_\mathrm{th}$ continuously for $\Delta T_\mathrm{PSB,IQ} = 40$~ms. However, it is allowed to accumulate time that $|\Delta\bar{I}_1'|>\Delta I_\mathrm{th}$ while the PSB activation signal is maintained. Using a pickup time of only 40~ms is possible thanks to the PSB activation function and results in enhanced dependability during fast and unconventional power swings. Moreover, to avoid the main PSB function from deactivating when $\bar{V}_1$ is low and $\bar{I}_1$ is high during subsequent power swings, a swing-in-progress logic is implemented, which maintains the PSB signal if it gets activated and $\bar{Z}_1$ crosses inside the supervisory zone. This is not a disadvantage compared to blinder-based PSB schemes which already have such a feature implemented since the PSB signal is maintained if $\bar{Z}_1$ crosses the blinders slow enough. Hence, the PSB unblocking for balanced faults occurring during power swings is a separate issue requiring a dedicated function (unbalanced faults can be detected using negative- and zero-sequence components). 

\subsection{The OST Function}
If $\bar{Z}_1$ is not inside the inner blinder in Fig.~\ref{fig:Blinder-example} when the IQ-based PSB activation signal is triggered, the OST function follows the same principle as explained in Section~\ref{sec:System-description}, meaning it stores in memory the time at which $\bar{Z}_1$ first enters, leaves and then enters the inner blinder a second time before issuing the OST command. However, if $\bar{Z}_1$ is inside the inner blinder when the IQ-based PSB activation signal is triggered, an OST signal might undesirable be triggered if the fault-to-swing trajectory of $\bar{Z}_1$ goes to the right in the impedance plane ($\bar{Z}_1$ leaves the inner blinder) followed by the IBR completing a full angular rotation before synchronizing ($\bar{Z}_1$ enters the inner blinder a second time). For the values of $\bar{Z}_\mathrm{s1}$ and $\delta$ in Fig.~\ref{fig:Z-traj-EC}~(a) and (b), together with the inner blinder in Fig.~\ref{fig:Blinder-example}, this situation may occur after reverse close-in or forward zone 2 faults are cleared. To avoid issuing the OST command for such stable power swings from IBRs, the OST logic is given an additional criterion which checks if $\bar{Z}_1$ is inside the inner blinder when the PSB signal is activated. If that is the case, the OST logic first requires $\bar{Z}_1$ to leave the inner blinder before being enabled.

\section{Simulation Results}
\label{sec:simulation-results}
In this section, simulation results on the system described in Section~\ref{sec:System-description} will be presented. The performance of the proposed IQ-based and conventional blinder-based PSB/OST schemes are compared. 

\subsection{Power Swings from SGs}
\begin{figure}[h]
    \centering
    \includegraphics[width=0.95\linewidth]{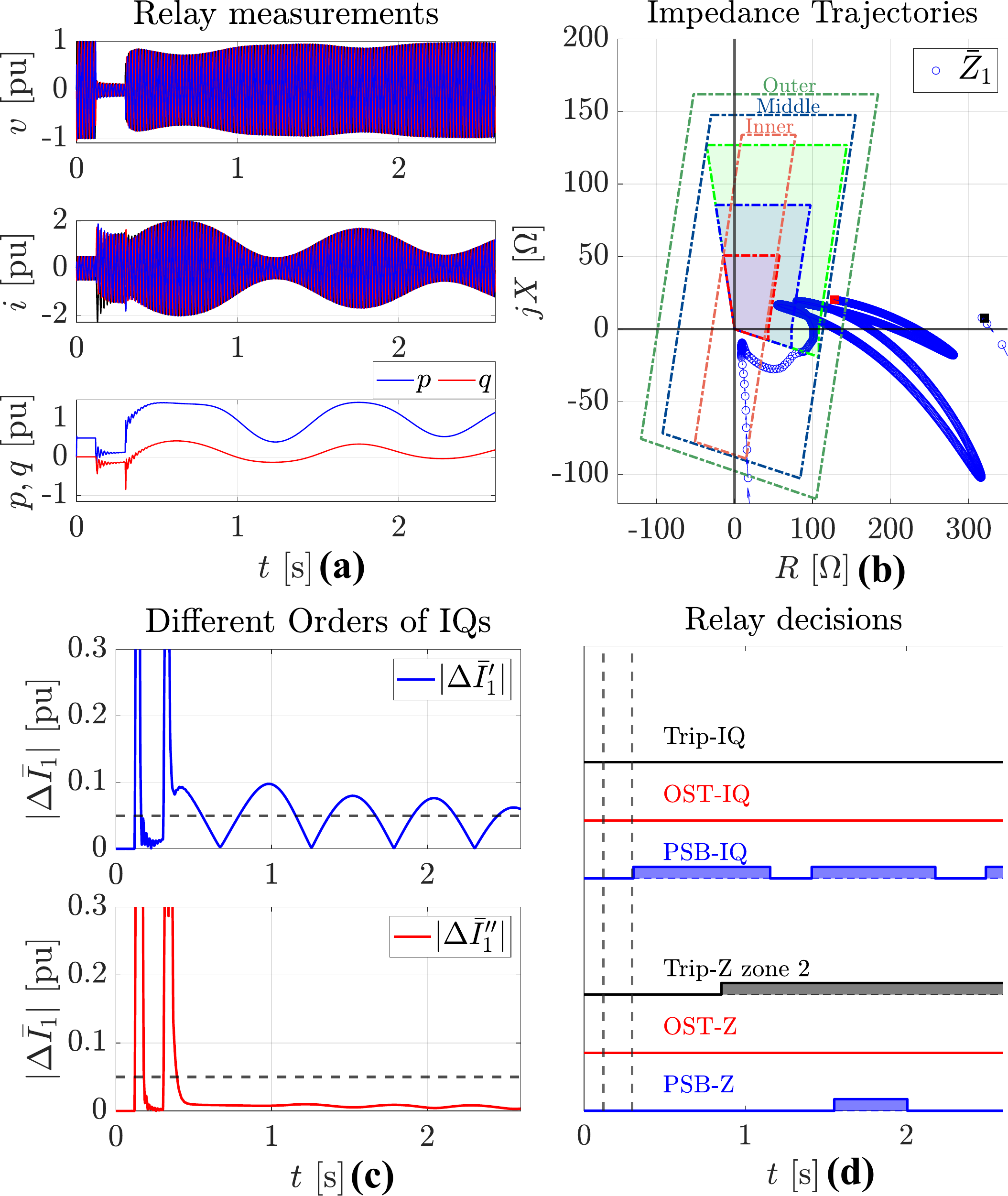}
    \caption{Barely stable swing with an SG after a close-in fault: \textbf{(a)} relay measurements (Rel1); \textbf{(b)} computed $\bar{Z}_1$; \textbf{(c)} computed IQs; \textbf{(d)} operation of PSB/OST schemes and distance zones.}
    \label{fig:ABCG_10_0_0_1_0.2_3_1SG_DPSP2027}
\end{figure}

Fig.~\ref{fig:ABCG_10_0_0_1_0.2_3_1SG_DPSP2027} shows the performance of the blinder- and IQ-based PSB/OST schemes at Rel1 for a stable power swing produced with an SG after a close-in fault in Fig.~\ref{fig:two-source-parallel-ohl}. The two dashed lines in Fig.~\ref{fig:ABCG_10_0_0_1_0.2_3_1SG_DPSP2027}~(d) mark the fault inception and clearance, respectively. It can be seen in Fig.~\ref{fig:ABCG_10_0_0_1_0.2_3_1SG_DPSP2027}~(b) that $\bar{Z}_1$ never leaves the blinders during the first swing cycle, causing distance zone 2 to trip with the blinder-based PSB scheme. As explained in Section~\ref{sec:SG-swings-explained}, such scenarios can occur with SGs, depending on the system impedances, the initial $\delta$ when the fault is cleared, the angular velocity $\delta'(t)$ and acceleration $\delta''(t)$. This is a limitation of blinders. However, the proposed IQ-based PSB scheme correctly manages to prevent the unwanted tripping.

\begin{figure}[h]
    \centering
    \includegraphics[width=0.95\linewidth]{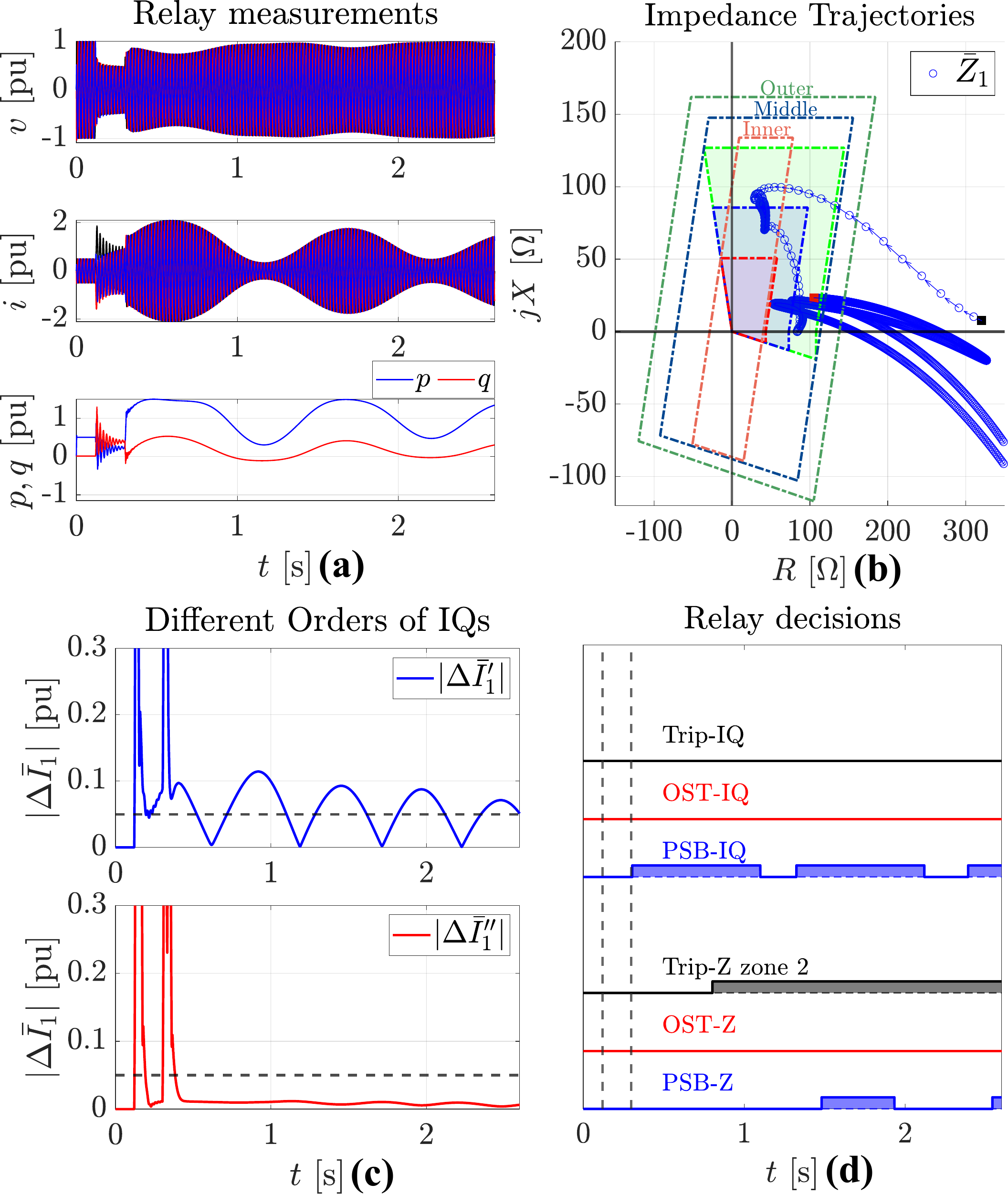}
    \caption{Barely stable swing with an SG.}
    \label{fig:ABCG_140_0_0_1_0.2_3_figs}
\end{figure}

It can be seen in Fig.~\ref{fig:ABCG_10_0_0_1_0.2_3_1SG_DPSP2027}~(c) that $\Delta I_\mathrm{th}=0.05$~pu and $\Delta T_\mathrm{PSB,IQ} = 60$~ms would be sufficient to prevent the PSB from activating during the close-in fault. On the other hand, if a zone 2 fault occurs instead (as seen from Rel1), these settings would not be adequate on their own, as shown in Fig.~\ref{fig:ABCG_140_0_0_1_0.2_3_figs}. However, with the proposed IQ-based PSB scheme which uses a PSB activation signal based on the second-order positive-sequence incremental current $|\Delta\bar{I}_\mathrm{1}''|$, the PSB scheme does not activate during the fault, allowing zone 2 to act as backup protection if the fault would last sufficiently long, or zone 1 to trip if $\bar{Z}_1$ eventually travels down into it.

If $\delta$ swings above $180^\circ$, the SG slips a magnetic pole and the power swing becomes unstable. In Fig.~\ref{fig:ABCG_10_0_0_1_0.2_3_1SG_DPSP2027}, this would cause $\bar{Z}_1$ to continue into distance zone 1 during the initial power oscillation, causing it to trip if the distance relay is equipped with the blinder-based PSB scheme. However, the proposed IQ-based PSB scheme prevents this from happening. The time it takes $\bar{Z}_1$ to enter distance zone 1 depends on the clearing angle $\delta$, and since the EC is located inside distance zone 1 with SGs, clearing the fault when $\delta$ is close to $180^\circ$ makes $\bar{Z}_1$ quickly enter distance zone 1 already during the fault-to-swing transition trajectory ($\bar{Z}_1$ goes straight up from the fault point). 

\begin{figure}[h]
    \centering
    \includegraphics[width=0.95\linewidth]{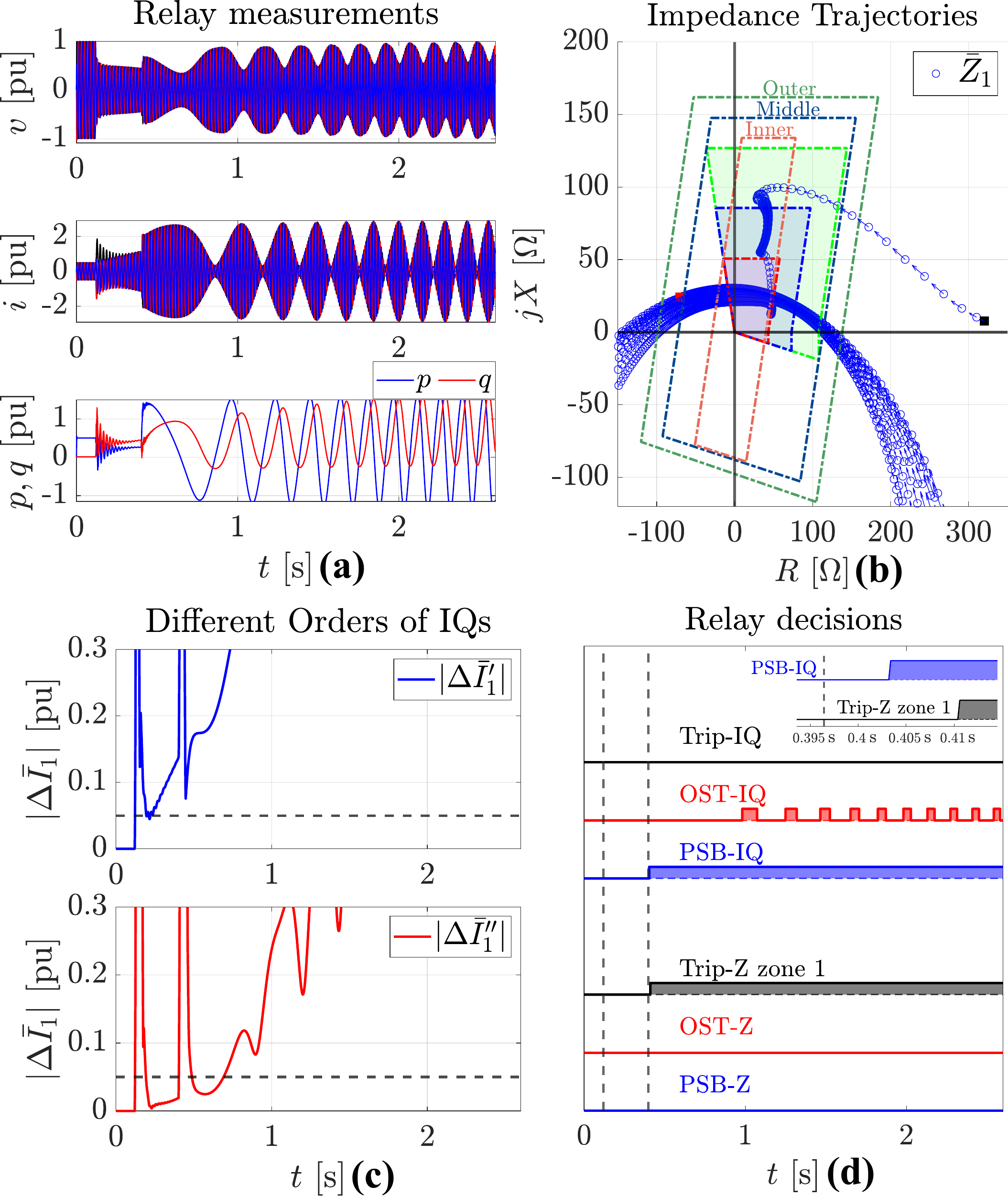}
    \caption{Unstable swing with an SG.}
    \label{fig:ABCG_140_0_0_1_0.3_3_figs}
\end{figure}

If an unstable power swing occurs after a zone 2 fault and is cleared when $\delta$ is close to $180^\circ$, the fault-to-swing transition trajectory of $\bar{Z}_1$ may move straight down from the fault point and into distance zone 1. Fig.~\ref{fig:ABCG_140_0_0_1_0.3_3_figs} shows such an unstable power swing observed in Rel1 when an SG is the sending-end source in Fig.~\ref{fig:two-source-parallel-ohl}. The fault on the parallel line is cleared after 300~ms and just in time before $\bar{Z}_1$ crosses into distance zone 1 during the fault itself. Since the faulted line has been cleared, it is crucial that the remaining healthy line does not also disconnect, which would cause the utility grid to suffer a substantial loss of power generation and experience a generation-load imbalance situation. As can be seen in Fig.~\ref{fig:ABCG_140_0_0_1_0.3_3_figs}, if Rel1 is equipped with the blinder-based PSB scheme, the healthy line trips during the $\bar{Z}_1$ fault-to-swing transition trajectory. On the other hand, the proposed IQ-based PSB scheme manages to block the distance element in time. The upper right corner of Fig.~\ref{fig:ABCG_140_0_0_1_0.3_3_figs}~(d) shows a zoomed-in plot of the time when the IQ-based PSB scheme asserts and when distance zone 1 issues a trip command if it is not blocked. It can also be seen that the continuous IQ-based PSB scheme does not disengage during the subsequent swing cycles due to $|\Delta\bar{I}_1|$ remaining far above the $\Delta I_\mathrm{th}=0.05$~pu threshold. Consequently, the OST is able to pick up on the subsequent swing cycles. The OST never picks up with the blinder-based PSB scheme because the initial pickup time $\Delta T_\mathrm{PSB,1}=30$~ms could not trigger on the first power oscillation, and for the subsequent swing cycles, the angular speed $\delta'(t)$ is already too fast, making $\bar{Z}_1$ cross between the outer and middle blinders faster than 30~ms.

\subsection{Power Swings from GFM Inverters}

\begin{figure}[h]
    \centering
    \includegraphics[width=0.95\linewidth]{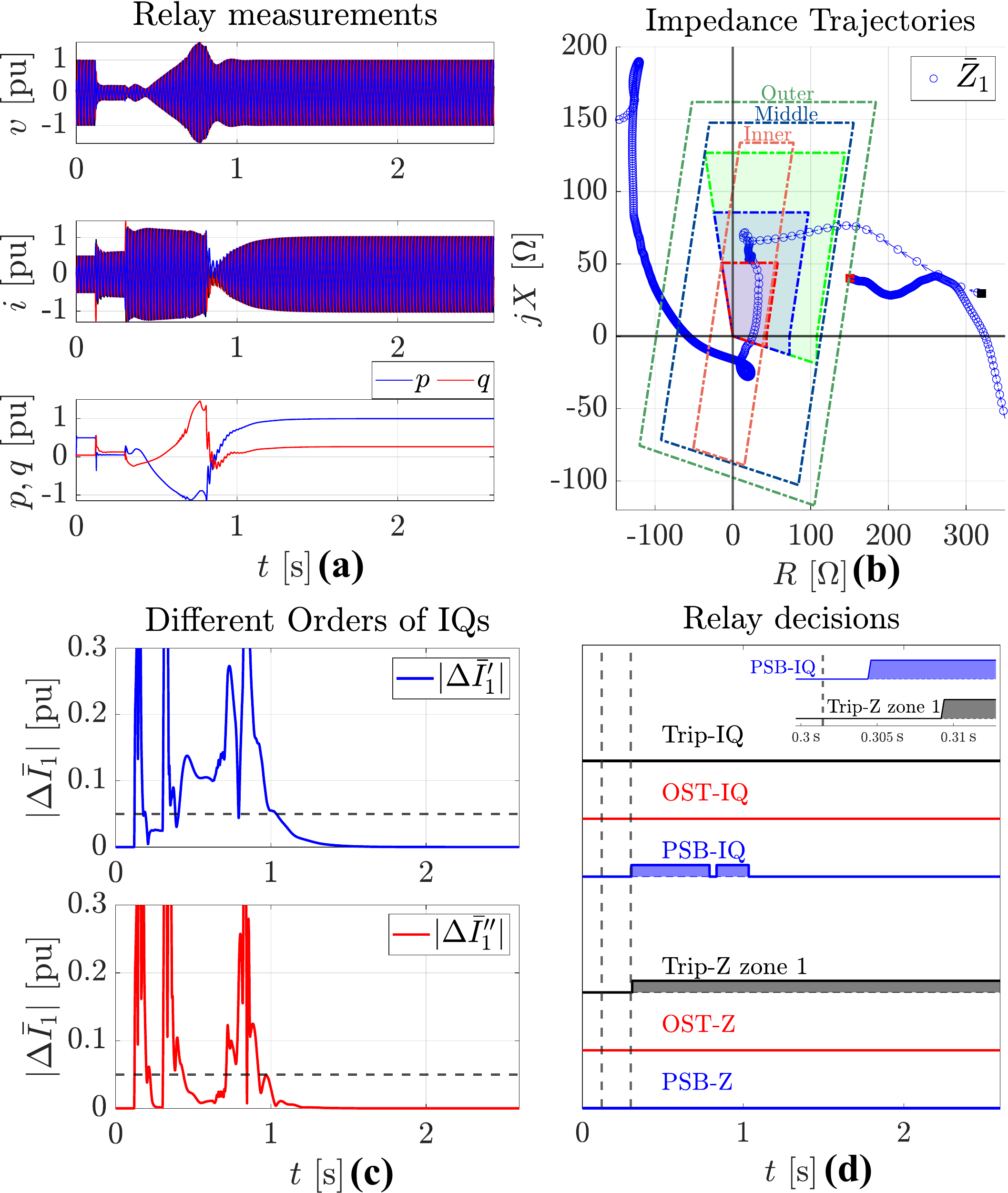}
    \caption{Stable swing with a droop-based GFM inverter.}
    \label{fig:ABCG_140_0_0_1_0.2_droop_figs}
\end{figure}

Fig.~\ref{fig:ABCG_140_0_0_1_0.2_droop_figs} shows the performance of the proposed IQ-based PSB scheme in Rel1 for a stable power swing produced with a droop-based GFM inverter after a zone 2 fault is cleared on the parallel line in Fig.~\ref{fig:two-source-parallel-ohl}. As can be seen, the $\bar{Z}_1$ fault-to-swing transition trajectory quickly enters distance zone 1, causing it to trip in less than 10~ms if the PSB scheme is based on conventional blinders. On the other hand, distance zone 1 is blocked before it has time to trip if the distance relay is equipped with the proposed IQ-based PSB scheme. A zoomed-in plot of its adequate operating speed is shown in the upper right corner of Fig.~\ref{fig:ABCG_140_0_0_1_0.2_droop_figs}~(d). Moreover, from Fig.~\ref{fig:ABCG_140_0_0_1_0.2_droop_figs}~(c) it is seen that $|\Delta \bar{I}_1'|$ is below $\Delta I_\mathrm{th}$ during the fault, which means that the main pickup time of $\Delta T_\mathrm{PSB,IQ}$ would be unable to prevent distance zone 1 from tripping during the $\bar{Z}_1$ fault-to-swing transition trajectory, which is another problem solved with the proposed PSB activation function using $|\Delta \bar{I}_1''|$. 

With regards to the $\bar{Z}_1$ trajectory during the power swing, it can be seen in Fig.~\ref{fig:ABCG_140_0_0_1_0.2_droop_figs}~(b) that $\bar{Z}_1$ travels below and then upwards but behind the distance zones, as explained in Section~\ref{sec:GFM-swings-explained} for GFM inverters emulating an adaptive virtual resistance. When $\delta=180^\circ$, the current limiter of the GFM inverter causes $\bar{Z}_\mathrm{s1}$ to be predominantly resistive, making the EC appear behind (to the left) of the distance zones. As $\delta$ continues beyond $180^\circ$, the current limiter is eventually disengaged and $\bar{Z}_1$ travels counterclockwise along a large circular arc until reaching the load region where the GFM inverter resynchronizes. Fig.~\ref{fig:ABCG_140_0_0_1_0.2_droop_figs}~(a) shows the first-order resynchronization dynamics of the active power (no overshoot, no oscillations) given by the droop-based PSC. Hence, even stable swings with droop-based GFM inverters may cause $\delta$ to traverse a complete turn and everything from relatively small to large clearing angles can cause problematic fault-to-swing transition trajectories of $\bar{Z}_1$ for blinder-based PSB schemes with GFM inverters. Lastly, it can be seen in Fig.~\ref{fig:ABCG_140_0_0_1_0.2_droop_figs}~(b) that the outer blinder is close to the new post-fault load region after one parallel line has been disconnected, which is a common design issue for blinders on heavily loaded lines \cite{PSB-Tutorial-Normann}. The proposed IQ-based PSB scheme does not need the outer blinder, mitigating the issue and allows for easier tuning on heavily loaded lines.

\begin{figure}[h]
    \centering
    \includegraphics[width=0.95\linewidth]{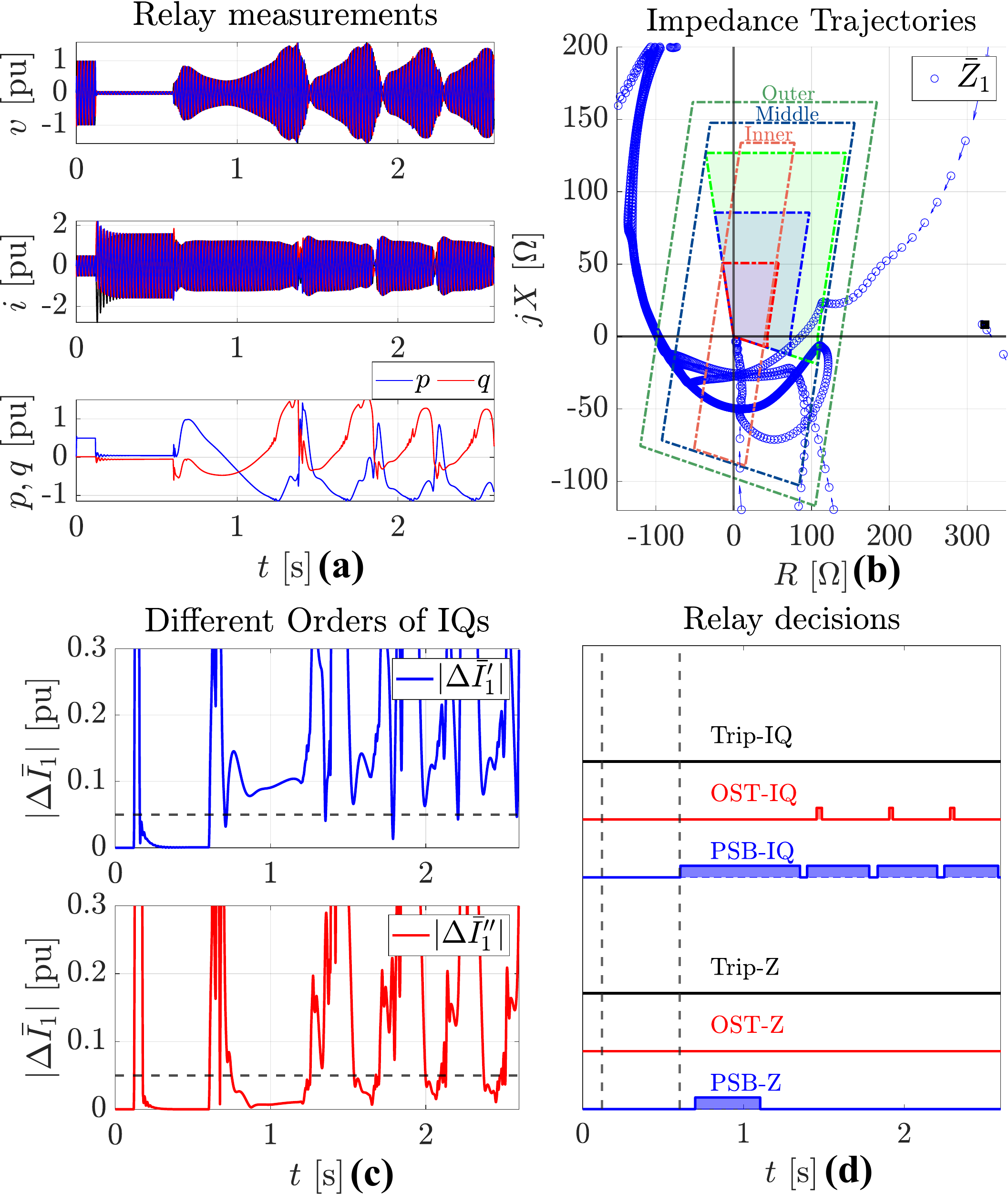}
    \caption{Unstable swing with a VSG-based GFM inverter.}
    \label{fig:ABCG_10_0_0_1_0.5_VSG_figs}
\end{figure}

The problematic $\bar{Z}_1$ fault-to-swing transition trajectories from zone 2 faults with GFM inverters having droop-based PSC also occur with VSG-based PSC. On the other hand, the VSG-based PSC can lose synchronism and cause unstable power swings. Because the GFM inverter in Fig.~\ref{fig:GFM1-vsg-droop} reduces its active power reference during faults, its angular advancement is slower than that of SGs. However, for sufficiently long-lasting faults, the VSG-based PSC may lose synchronism. Fig.~\ref{fig:ABCG_10_0_0_1_0.5_VSG_figs} shows such a scenario from a close-in balanced fault on the parallel line in Fig.~\ref{fig:two-source-parallel-ohl}. The plotted current in Fig.~\ref{fig:ABCG_10_0_0_1_0.5_VSG_figs}~(a) is the total current measured by Rel1, which is higher than 1.2~pu during the fault, but the injected current from the GFM inverter is strictly limited to 1.2~pu.

It can be observed in Fig.~\ref{fig:ABCG_10_0_0_1_0.5_VSG_figs} that the swing frequency is not high---it is indeed slower than in Fig.~\ref{fig:ABCG_140_0_0_1_0.3_3_figs} with the SG. However, for the three plotted unstable swing cycles, $\bar{Z}_1$ accelerates precisely when crossing the blinders. It only takes $\bar{Z}_1$ 3-5~ms to cross between the outer and middle blinders at the lower-right corner and around 9~ms when crossing in the first quadrant. This is mainly caused by the GFM inverter current limiter quickly increasing the magnitude and changing the angle of $\bar{Z}_\mathrm{s1}$ at the start of each subsequent swing cycle, which moves the EC location and the value of $\bar{Z}_1$ along with it, but also that it switches between active and reactive power prioritization depending on the PCC voltage level by adjusting the active power reference $P^*$. Hence, despite the blinder-based PSB scheme having a pickup time of $\Delta T_\mathrm{PSB,2} = 10$~ms after detecting the first oscillation, it does not activate on the subsequent swing cycles, and consequently the OST never activates. The proposed IQ-based PSB/OST scheme operates correctly. 

\section{Conclusion}
\label{sec:conclusions}
This paper proposes an IQ-based PSB scheme for distance relays. The standard first-order incremental currents are used for the main PSB function, but an additional PSB activation function is implemented based on higher orders of incremental currents. The proposed IQ-based PSB scheme is able to handle fast movements of the apparent impedance caused by unconventional power swings from IBRs that blinder-based PSB schemes struggle to pickup. It also reliably prevents the distance zones from tripping during the fault-to-swing transition trajectory of the apparent impedance and the initial power oscillation when blinder-based PSB schemes cannot operate. It is shown that the varying source impedance of GFM inverters can make such difficult swing scenarios occur for a wider range of fault clearing angles under both stable and unstable power swings. The superior performance of the proposed IQ-based PSB scheme over traditional blinder-based schemes is validated with PSCAD/EMTDC simulations of an SG and GFM inverter implementation.

\section{Acknowledgement}
The authors thank Hitachi Energy and the Swedish Energy Agency who have financially supported this work.


\section{References}
\vspace*{6pt}


\begin{thebibliography}{14}

\bibitem{PSB-Tutorial-Normann} Fischer, N., Benmouyal, G., Hou, D., et al.: `Tutorial on power swing blocking and out-of-step tripping'. Presented at the 39th Annual Western Protective Relay Conference, Spokane, Washington, 2004

\bibitem{PSB-OOS-fundamentals} Tziouvaras, D.A., and Hou, D.: 'Out-of-step protection fundamentals and advancements', in IEEE 57th Annual Conference for Protective Relay Engineers, 2004, pp. 282–307

\bibitem{PSB-Bogdan-Zcont} Kasztenny, B., West, M., Mynam, V. M.: `Power-swing detection logic for systems with low inertia'. Proc. 17th International Conference on Developments in Power System Protection (DPSP 2024), 2024, vol. 2024, pp. 132–141

\bibitem{REL670-application-manual} Hitachi Energy: `Line distance protection REL670 Version 2.2 ANSI—Application manual', 2025, \url{https://www.hitachienergy.com/products-and-solutions/substation-automation-protection-and-control/products/protection-and-control/relion-product-family/relion-670-series/rel670}


\bibitem{D60-instruction-manual-GE} General Electric: `D60 Line Distance Protection System—Instruction Manual', 2022, \url{https://www.gevernova.com/electrification/sites/default/files/resources/products/manuals/d60/D60-1601-0089-79x-AI3-GEV.pdf}

\bibitem{SEL-421-4} Schweitzer Engineering Laboratories: `SEL-421-4, -5 Protection, Automation, and Control System—High-Speed Line Protection, Automation, and Control System', 2025, \url{https://selinc.com/products/421/docs/}

\bibitem{CIGRE-Protection-Challenges-IBRs} Nagpal, M., Jensen, M., Higginson, M., et al.: `Protection challenges and practices for interconnecting inverter based resources to utility transmission systems', 2020

\bibitem{PSB-impact-of-wind} Haddadi, A., Kocar, I., Karaagac, U., et al.: `Impact of wind generation on power swing protection', IEEE Transactions on Power Delivery, 2019, 34, (3), pp. 1118–1128

\bibitem{PSB-IBRs-PartII} Nasr, M.-A., Hooshyar, A.: `Power swing in systems with inverter-based resources—part ii: Impact on protection systems', IEEE Transactions on Power Delivery, 2024, 39, (3), pp. 1903–1917

\bibitem{PSB-GFM-VI} Niu, Y., Yang, Z., Pal, B. C.: `Power swing trajectory influenced by virtual impedance-based current-limiting strategy', IEEE Transactions on Smart Grid, 2026, 17, (1), pp. 415–428

\bibitem{PSB-Yongxin-GFL} Xiong, Y., Wu, H., Li, Y., et al.: `Comparison of power swing characteristics and efficacy analysis of impedance-based detections in synchronous generators and grid-following systems', IEEE Transactions on Power Systems, 2025, 40, (3), pp. 2545–2556

\bibitem{PSB-Yongxin-GFM} Xiong, Y., Wu, H., Li, Y., et al.: `Analysis of power swing characteristics of grid-forming VSC system considering the current limitation mode', IEEE Transactions on Power Systems, 2026, pp. 1–12

\bibitem{PSB-GFM-Protection-and-Dynamics} Chen, X., Haddadi, A., Farantatos, E., et al.: `Power swing in systems with varying penetration of grid-forming IBRs: Protection and dynamics', Protection and Control of Modern Power Systems, 2025, 10, (4), pp. 116–129

\bibitem{HJ-GFM-distance} Johansson, H., Xing, Q., Taylor, N., et al.: `Impacts of grid-forming inverters on distance protection', IET Generation, Transmission \& Distribution, 2025, 19, (1), p. e13354. Available at: wiley.com

\bibitem{GFM-limiter-sat-resistive} Fan, B., Wang, X.: `Equivalent circuit model of grid-forming converters with circular current limiter for transient stability analysis', IEEE Transactions on Power Systems, 2022, 37, (4), pp. 3141–3144


\end{thebibliography}
\end{document}